\documentclass[%
aps,%
prd,%
preprintnumbers,%
twocolumn,%
prd,
nofootinbib]{revtex4}

\usepackage{graphicx}
\usepackage{amsmath}

\renewcommand{\ol}{\overline}

\newcommand{\wt}{\widetilde}

\newcommand{\eqn}[1]{&\hspace{-0.3em}#1\hspace{-0.3em}&}

\newcommand{\GeV}{\mbox{GeV}}
\newcommand{\TeV}{\mbox{TeV}}
\begin{document}

\preprint{TU-768}

\title{Gravitinos from Heavy Scalar Decay}

\author{Takehiko Asaka, Shuntaro Nakamura and Masahiro Yamaguchi}

\affiliation{
Department of Physics, Tohoku University, Sendai 980-8578, Japan
}

\date{\today}

\begin{abstract}
    Cosmological issues of the gravitino production by the decay of a
    heavy scalar field $X$ are examined, assuming that the damped
    coherent oscillation of the scalar once dominates the energy of
    the universe. The coupling of the scalar field to a gravitino pair
    is estimated both in spontaneous and explicit supersymmetry
    breaking scenarios, with the result that it is proportional to the vacuum
    expectation value of the scalar field in general. Cosmological
    constraints depend on whether the gravitino is stable or not, and
    we study each case separately.  For the unstable gravitino with
    $M_{3/2} \sim$ 100GeV--10TeV, we obtain not only the upper bound,
    but also the lower bound on the reheating temperature after the
    $X$ decay, in order to retain the success of the big-bang
    nucleosynthesis. It is also shown that it severely constrains the
    decay rate into the gravitino pair.  For the stable gravitino,
    similar but less stringent bounds are obtained to escape the
    overclosure by the gravitinos produced at the $X$ decay. The
    requirement that the free-streaming effect of such gravitinos
    should not suppress the cosmic structures at small scales
    eliminates some regions in the parameter space, but still leaves a
    new window of the gravitino warm dark matter.  Implications of
    these results to inflation models are discussed.  In particular,
    it is shown that modular inflation will face serious cosmological
    difficulty when the gravitino is unstable, whereas it can escape
    the constraints for the stable gravitino.  A similar argument
    offers a solution to the cosmological moduli problem, in which the
    moduli is relatively heavy while the gravitino is light.
\end{abstract}


\maketitle
%
\section{Introduction}
It is quite plausible that the universe once experienced the epoch
where its energy density is dominated by the coherent oscillation of a
scalar field. A typical example is the inflaton oscillation after the
exit of the de Sitter expansion period in the slow roll
inflation~\cite{Linde:1981mu}.  Another example is dilaton and moduli
fields in superstring theories.

The oscillating field eventually decays, followed by the reheating of
the universe.  Then, the radiation dominated era (the hot big-bang
universe) commences. At the same time, however, the decay of the
oscillating field may produce unwanted particles.%
\footnote{
In general, the moduli fields lead to cosmological difficulties
known as the moduli problem~\cite{Coughlan:1983ci}.}
Recently, it has
been recognized that the gravitino production at the decay of a
modulus field can cause cosmological
disasters~\cite{Endo:2006zj,Nakamura:2006uc}. It turns out that the
decay width into the gravitino pair is non-suppressed
 and is
comparable to that into other particles which couple to the moduli
field with Planck suppressed interaction. Thus, the branching ratio of
the moduli decay into the gravitino pair is sizable: in fact it is
typically at a percentage level, or even higher. If the damped
coherent oscillation of the modulus field once dominates the energy
density of the universe, the overproduction of the gravitinos at its
decay would cause serious problems to cosmology. The gravitino decay,
if it is unstable, would spoil the success of the big-bang
nucleosynthesis (BBN). In addition, these unstable gravitinos decay into the
lighter superparticles, resulting in the overabundance of the lightest
superparticles (LSPs). These arguments push up the gravitino mass to a region
disfavored from the naturalness problem associated with the weak
scale~\cite{Nakamura:2006uc}.

One should be aware that the problem of the gravitino overproduction may apply
to other heavy scalar fields. In particular, the case of inflaton is
important. After the epoch of de Sitter expansion driven by the vacuum
energy, the inflaton field falls down to its true minimum and starts
the damped coherent oscillation around the true minimum. Eventually it
decays to reheat the universe.  In many inflation models, the mass of
the inflaton, or more precisely the mass of the oscillating field
after the slow-roll inflation, is much larger than the weak
scale. Thus,  in the low energy supersymmetry, the decay into a gravitino
pair is likely to be kinematically allowed. If the branching ratio is
not negligible, then the gravitino production at the inflaton decay
will cause serious cosmological problems.

In this paper, we would like to investigate the decay of a heavy
scalar field into a gravitino pair in a general ground and consider
its cosmological implications.  In Sec. II, we shall first discuss the
partial decay rate of the scalar decay into the gravitino pair.  The
decay amplitude is proportional to the vacuum expectation value (VEV)
of the $F$ auxiliary component of the scalar field. We will estimate
the VEV of the $F$ and thus the decay width into the gravitino pair in
a very general setting.  Then, in Sec. III, we will discuss
cosmological problems caused by the gravitino problem.  We will carry
out the study in both unstable and stable gravitino cases, and
identify the region of the parameter space which survives various
cosmological constraints. A particular attention is paid to the case
where the scalar field shares the properties with the moduli fields.
We will show that the case faces a serious cosmological difficulty
when the gravitino is unstable, but can survive the constraints for a
light and stable gravitino. We will also point out a new window of the
gravitino warm dark matter with the mass range of 10 MeV to 1 GeV.
In Sec. IV, we will draw our conclusions and also discuss 
 implications of our results to the
inflation model building.  In Appendix, we will 
explain the details on the estimation of the abundance for
the long-lived superparticles.

\

While preparing the paper, we received a
preprint~\cite{Kawasaki:2006gs}, which dealt with a similar subject.
Our results agree with Ref.~\cite{Kawasaki:2006gs}, where overlap.
\section{Heavy scalar decay into gravitinos}
Let us begin by discussing the decay of a heavy scalar field $X$
into a pair of gravitinos.  To avoid unnecessary complication, we consider 
the case where the mass of $X$ is much heavier than the gravitino mass, $M_X \gg
M_{3/2}$.  Also we assume that $X$ is a singlet under the standard model
gauge group. These assumptions are relevant for later use.

The scalar field $X$ may decay into a pair of gravitinos:
\begin{eqnarray}
    \label{eq:Xto32}
    X \to \psi_{3/2} + \psi_{3/2} \,,
\end{eqnarray}
The decay is induced through the interaction in the supergravity.
As recently calculated in Refs.~\cite{Endo:2006zj,Nakamura:2006uc},
its partial decay width is given by
\begin{eqnarray}
    \label{eq:G32}
    \Gamma_{3/2} = \frac{d_{3/2}^2}{288 \pi} \frac{M_X^3}{M_P^2} \,,
\end{eqnarray}
where $M_P \simeq 2.4 \times 10^{18}$ GeV is the reduced Planck scale.
The coupling constant $d_{3/2}$ is defined by the relation
\begin{eqnarray}
    d_{3/2} \frac{M_{3/2}^2}{M_X} 
    = 
    \left\langle \left( G_{X \ol X} \right)^{- \frac 1 2} e^{G/2} G_X 
    \right\rangle
    \,,
\end{eqnarray}
where $G$ is the total K\"ahler potential, $G \equiv K + \ln |W|^2$,
with $K$ and $W$ being the K\"ahler potential and the superpotential,
respectively. The subscript $X$ ($\ol X$) denotes differentiation with
respect to the $X$ ($\ol X$) field and $\langle \cdots \rangle$ stands
for the VEV. Note that the right-handed
side of the above equation is the (canonically normalized)
$F$-auxiliary component of the $X$ field. Therefore, $d_{3/2}$
parameterizes the supersymmetry breaking felt by the $X$ field.  The
parameter $d_{3/2}$ is expected to be order unity for a moduli field.
If this is the case, the decay into the gravitino pair can be
significant, in particular when the decays into other particles are
also mediated by Planck suppressed interactions. Then, gravitinos
produced at the scalar decay would lead to cosmological disasters.  A
more general case with less gravitino production will be discussed
later.

To proceed, we now present a general expression for the VEV of the
$X$'s auxiliary field.%
\footnote{ A similar analysis can be found in
  Ref.~\cite{Joichi:1994tq}.  }
In the absence of $D$-terms, the scalar
potential of the ${\cal N}=1$ supergravity is given
\begin{eqnarray}
   V_{\rm SUGRA}=e^G \left(G_i G^{i \bar{j}} G_{\bar{j}}-3 \right),
\label{eq:SUGRA}
\end{eqnarray}
where the subscript $i$ ($\bar{i}$) represents the derivative with respect
to a scalar field $\phi^i$ ($\ol \phi^{\bar{i}}$) and  
$G^{i \bar{j}}$ is the inverse of the K\"ahler metric. 
The first derivative of the potential is obtained as
\begin{eqnarray}
    &&\frac{\partial}{\partial \phi^i} V_{\rm SUGRA}
    \nonumber \\
    \eqn{=} G_i V_{\rm SUGRA}
    \nonumber \\
    \eqn{}+e^G \left( G_{ij} G^{j \bar{k}} G_{\bar{k}} 
        + G_j G^{j \bar{k}}{}_{i} G_{\bar{k}} 
        + G_j G^{j \bar{k}} G_{\bar{k} i} 
    \right)
    \nonumber \\
    \eqn{=}
    G_i V_{\rm SUGRA}
    \nonumber \\
    \eqn{}+e^G \left( G_{ij} G^{j \bar{k}} G_{\bar{k}} 
        - G_j G^{j \bar{l}} G_{\bar{l} m i}G^{m \bar{k}} G_{\bar{k}}
        +G_i \right). 
\end{eqnarray}
Noticing that the $F$ auxiliary field is expressed, in terms of $G$ and its
derivatives, as 
\begin{eqnarray}
   F^i=-G^{i \bar{j}} e^{G/2} G_{\bar{j}}, 
\end{eqnarray}
we find
\begin{eqnarray} 
 \frac{\partial }{\partial \phi^i}  V_{\rm SUGRA}
& =& -e^{G/2} G_{ij}F^j -e^{G/2}(1+\tilde{V})G_{i \bar{j}} \ol F^{ \bar{j}}
\nonumber \\
& &       -G_{ij \bar{k}} F^j \ol F^{\bar{k}} \,,
\end{eqnarray}
with $\widetilde{V} \equiv G_i G^{i \bar{j}} G_{\bar{j}}-3$. 

Consider first the case where the supersymmetry is spontaneously
broken within the framework of supergravity. In this case, in addition
to the $X$ field, we need another field $Z$ which dominantly breaks
supersymmetry.  At the vacuum, the VEV of the K\"ahler metric can be
canonically normalized as $\langle G_{i \bar{j}} \rangle =
\delta_{i \bar{j}}$. With this basis, the stationary point condition reads
\begin{eqnarray}
  0&=&  \left \langle \frac{\partial}{\partial X} V_{\rm SUGRA} \right \rangle
\nonumber \\
   &=& -\left\langle e^{G/2} G_{XX} \right\rangle \langle F^X \rangle 
    -\left\langle e^{G/2} G_{XZ} \right\rangle \langle F^Z\rangle
\nonumber \\
   & &  
    -\left\langle e^{G/2} (1 + \widetilde{V}) \right\rangle 
     \langle \ol F{}^{\bar{X}}      \rangle
    -\langle G_{X Z \bar{Z}} \rangle \langle F^Z \rangle 
                                    \langle \ol F{}^{\bar{Z}} \rangle
\nonumber \\
   & &  
   - \langle G_{X X \ol Z} 
   \rangle \langle F^X \rangle
   \langle \ol F{}^{\ol Z} \rangle
   - \langle G_{X X \ol X} \rangle 
   \langle F^X \rangle
   \langle \ol F{}^{\ol X} \rangle
\nonumber \\
   & &  
   - \langle G_{X Z \ol X} \rangle \langle F^Z \rangle
   \langle \ol F{}^{\ol X} \rangle \,.
\label{eq:stationary-point-condition}
\end{eqnarray}
Notice that 
\begin{eqnarray}
    \langle e^{G/2} \rangle =  M_{3/2}\,,~~~  
    \langle F^Z \rangle \simeq M_{3/2}.
\label{eq:gravitino-mass}
\end{eqnarray}
When the mass of the $X$ field is much heavier than the
gravitino mass, one finds
\begin{eqnarray}
\left\langle e^{G/2} G_{XX} \right\rangle
&=&\left\langle e^{G/2} \left( K_{XX}-\frac{W_X^2}{W^2}+\frac{W_{XX}}{W}
\right) \right\rangle
\nonumber \\
&\simeq& 
\left\langle e^{G/2} \left( K_{XX}-K_X^2+\frac{W_{XX}}{W} 
\right) \right\rangle
\nonumber \\
&\simeq& \left\langle e^{K/2} W_{XX} \right\rangle +{\cal O}(M_{3/2})
\nonumber \\
&\simeq& M_X +{\cal O}(M_{3/2}) \,.
\label{eq:X-mass}
\end{eqnarray}
In the above we have used $\langle K_X+W_X/W \rangle \ll 1$.

Using Eqs.~(\ref{eq:stationary-point-condition}), (\ref{eq:gravitino-mass})
and (\ref{eq:X-mass}), the VEV of the $F$-auxiliary field can be approximately
expressed as
\begin{eqnarray}
    \label{eq:FVEVX}
    \langle F^X \rangle 
    &\simeq & 
    -\frac{1}{M_X} \Bigl(
    \langle G_{XZ} \rangle M_{3/2} \langle F^Z \rangle
    \nonumber \\ 
    & & ~~~~~~~~~ + 
    \langle G_{X Z \bar{Z}} \rangle \langle F^Z \rangle 
    \langle \ol F{}^{\bar{Z}} \rangle \Bigr) \,.
\label{eq:expression-F_X}
\end{eqnarray}
Here, evaluation of $\langle  G_{XZ} \rangle$ can be done as follows:
\begin{eqnarray}
 \left\langle  G_{XZ} \right\rangle
 &  =& \left\langle   K_{XZ}+\frac{W_{XZ}}{W}
                           -\frac{W_X}{W} \frac{W_Z}{W} \right\rangle 
\nonumber \\ 
& \simeq &
 \left\langle  K_{XZ}+\frac{W_{XZ}}{W}
                           +K_X \frac{W_Z}{W}  \right\rangle \,.
\end{eqnarray}
It is natural that  $K_X $ and $ K_{XZ} $ have VEVs comparable to 
$\langle X \rangle$.%
\footnote{ This is the case when the K\"ahler potential takes the form
  of $K = f(Z,\ol Z) X \ol X + \cdots$ with $\langle Z \rangle = {\cal
    O}(1)$, for example.  }
Furthermore, $\langle W_Z/W \rangle ={\cal O} (1)$.  Assuming that
$\langle W_{XZ}/ W \rangle$ is negligibly small, which is the case
when the $X$ field is separated from the supersymmetry breaking sector
in the superpotential, we find
\begin{eqnarray}
    \langle  G_{XZ} \rangle ={\cal O}( \langle X \rangle ).
\end{eqnarray}
On the other hand, we also expect that $\langle G_{X Z \bar{Z}} \rangle 
= \langle K_{X Z \bar{Z}} \rangle \simeq {\cal O}(\langle X \rangle )$.
To summarize the evaluations given above, we conclude%
\footnote{
For a moduli field, $\langle X \rangle = {\cal O}(1)$. The relation should
simply read
$|\langle F^X \rangle | \simeq M_{3/2}^2/M_X$.}
\begin{eqnarray}
   \left| \langle F^X \rangle \right| 
   \simeq 
   \frac{M_{3/2}^2}{M_X} 
   \left| \langle X \rangle \right|,
\label{eq:F_X}
\end{eqnarray}
as long as there is no cancellation between the terms
in the right-handed side of Eq.~(\ref{eq:FVEVX}).
This in turn implies 
\begin{eqnarray}
    d_{3/2} \simeq 
    \left| \langle X \rangle \right| \,,
\end{eqnarray}
in the Planck unit.%
\footnote{ For example, in the case where $W = W(X) + \Lambda^2 ( Z +
  \beta )$~\cite{Polonyi:1977pj} and the minimal K\"ahler potential,
  we find that $d_{3/2} = \sqrt{3} |\langle X \rangle|$ in the Planck
  unit.  }

Next, we would like to discuss the case where the supersymmetric
anti-de Sitter vacuum is uplifted to the Minkowski one by explicit
SUSY breaking terms. This is indeed the case for the KKLT
set-up~\cite{Kachru:2003aw}, where the explicit SUSY breaking is
originated from an anti-D3 brane. The scalar potential of the
effective theory is then written in the form
\begin{eqnarray}
    V=V_{\rm SUGRA} + V_{\rm breaking} \,,
\label{eq:explicit-breaking}
\end{eqnarray}
where $V_{\rm SUGRA}$ is the scalar potential of the supergravity given in
Eq.~(\ref{eq:SUGRA}). $V_{\rm breaking}$ is the explicit SUSY breaking term
which is of the form 
\begin{eqnarray}
    V_{\rm breaking}= \hat{V} f(X, \ol X) \,,
\end{eqnarray}
with some real function $f(X, \ol X)$.  

The stationary point condition of the scalar potential 
(\ref{eq:explicit-breaking}) with respect to $X$ now reads
\begin{eqnarray}
  \left\langle \frac{\partial V}{\partial X} \right\rangle
& =& \hat{V} \left\langle \frac{\partial f}{\partial X} \right\rangle
   +\left\langle \frac{\partial}{\partial X} V_{\rm SUGRA} \right\rangle
\nonumber \\
& \simeq & \hat{V} \left\langle \frac{\partial f}{\partial X} \right\rangle
    -M_X \langle F^X \rangle 
\nonumber \\
& =& 0.
\end{eqnarray}
It follows from the above that
\begin{eqnarray}
   \langle F^X \rangle \simeq \frac{1}{M_X} \hat{V} 
     \left\langle \frac{\partial f}{\partial X} \right \rangle.
\end{eqnarray}
By imposing that the vacuum
energy vanishes at the minimum, we can determine the value of 
$\hat{V}$:  $\hat{V}=3 M_{3/2}^2/\langle f \rangle$. Then we obtain
\begin{eqnarray}
   \langle F^X \rangle =\frac{3 M_{3/2}^2}{M_X} 
      \left\langle \frac{\partial}{\partial X} \ln f \right\rangle.
\end{eqnarray}
To proceed further, let us see a set-up of
KKLT-type~\cite{Kachru:2003aw}. The function $f$ is
likely of the form~\cite{Kachru:2003sx,Witten:1985xb}
\begin{eqnarray}
   \ln f= -n \ln ( T + \ol T - \ol C C) \,,
\end{eqnarray}
where $T$ is an over-all modulus field and $C$ represents a (matter)
field, with $n$ being some constant of order unity.  With this form,
we obtain 
$\langle \partial \ln f /\partial T \rangle = {\cal O}(\langle T \rangle^{-1})
={\cal O}(1)$ and 
$\langle \partial \ln f /\partial C \rangle \simeq {\cal O} (\langle C
\rangle )$, which implies $\langle \partial \ln f /\partial X \rangle
\simeq {\cal O} (\langle X \rangle )$. Thus, we reproduce the same
relation (\ref{eq:F_X}) even when supersymmetry is explicitly 
broken.\footnote{
Estimation for an overall moduli case was given in Ref.~\cite{Choi:2005ge}.
}

It is amusing to point out that the relation (\ref{eq:F_X}) can be
understood in softly broken global SUSY. 
For example, let us consider the following Lagrangian:
  \begin{eqnarray}
    \mathcal{L} &=& \Bigl[ X^{\dagger} X + \left( \theta^2 M_1 X^{\dagger} X + \mathrm{h.c.} \right) 
       - \theta^2 \bar{\theta}^2 M_2^2 X^{\dagger} X 
                  \Bigr]_D
\nonumber \\
& &                   + \left[ W(X) \right]_F + \mathrm{h.c.}, 
  \end{eqnarray}
with  $M_1$ and $M_2$  the soft SUSY breaking parameters  of order $M_{3/2}$.
Here $X$ denotes a chiral supermultiplet and $X^{\dagger}$ its hermitian
 conjugate, unlike 
the rest of the paper. We can eliminate the auxiliary field $F^X$ by using the equations of motion
\begin{eqnarray}
     F^X=- \left(\overline{W}_{\overline{X}}+M_1 X \right).
\end{eqnarray}
The resulting scalar potential is written
  \begin{equation} \label{eq: sca_pote}
    V = \left| M_1 \overline{X} + W_X \right|^2 + M_2^2 \overline{X} X. 
  \end{equation}
The stationary condition leads
  \begin{align}
    0 &= \bigg\langle \frac{\partial V}{\partial X} \bigg\rangle \nonumber \\
      &= \Big\langle M_1 X + \overline{W}_{\bar{X}} \Big\rangle \langle W_{XX} \rangle 
         + M_1 \Big\langle M_1 \overline{X} + W_X \Big\rangle 
         + M_2^2 \langle X \rangle \nonumber \\ 
      &\simeq -M_X \langle F^X \rangle + M_2^2 \langle X \rangle.
  \end{align}
Thus, we obtain 
\begin{equation}
    \left| \langle F^X \rangle \right| 
    = \frac{M_2^2}{M_X} 
    \left| \langle X \rangle \right| \,,
\end{equation}
which is similar to Eq.~(\ref{eq:F_X}). This argument implies our
result is insensitive whether the origin of SUSY breaking is
spontaneous or explicit.

\section{Cosmology of gravitinos}
%
\subsection{Gravitino abundance}
Let us discuss the cosmological implications of gravitinos
which are produced by the heavy scalar field $X$. 
We will consider the case in which 
$X$ dominates the energy of the universe when it decays.
This situation can be achieved if $X$ obeys the coherent oscillation 
with a large initial amplitude and also it is long-lived.
In this case, the decay of $X$ reheats the universe.  We define
here the reheating temperature $T_R$ by
\begin{eqnarray}
    \label{eq:defTR}
    T_R \equiv 
    \left( \frac{90}{\pi^2 g_\ast (T_R)} \right)^{\frac 14} 
    \sqrt{ \Gamma_X M_P } \,,
\end{eqnarray}
where $g_\ast (T_R)$ is the number of the effective degrees of freedom
at $T = T_R$, and the total decay rate of $X$ is denoted by
$\Gamma_X$.  Notice that $\Gamma_X$ is determined by the interactions
of $X$ to other light particles in the minimal supersymmetric standard
model (MSSM), and so it is highly model dependent.   
To parameterize the ignorance of the strength of interaction, 
we introduce $d_{\rm tot}$  to express the total decay rate
\begin{eqnarray}
    \label{eq:GAMX_MODULI}
    \Gamma_X = \frac{d_{\rm tot}^2}{8 \pi} \frac{M_X^3}{M_P^2} \,.
\end{eqnarray}
When $X$ has only Planck suppressed interaction,  $d_{\rm tot}$ becomes  
of order unity, as far as the Born approximation is valid.  For example, when
$X$ is the  heavy moduli field and couples to gauge supermultiplets through
Planck suppressed interaction in the gauge kinetic function, 
the decay rate of $X$ is computed to be in the form
Eq.~(\ref{eq:GAMX_MODULI}) with $d_{\rm tot} = {\cal
  O}(1)$~\cite{Endo:2006zj,Nakamura:2006uc}.  With Eq.~(\ref{eq:GAMX_MODULI}),
the reheating temperature is estimated as
\begin{eqnarray}
    \label{eq:TRMODULI}
    T_R = 5.9 \times 10^4 \GeV  \, d_{\rm tot} \,
    \left( \frac{M_X}{10^{10}\GeV}\right)^{\frac 32} \,,
\end{eqnarray}
where we have used $g_\ast (T_R) = 200$.
It should be noted that the reheating temperature should be
$T_R \gtrsim 7$ MeV to be consistent with the BBN
theory~\cite{Kawasaki:1999na}, which
means that the mass of $X$ should be 
\begin{eqnarray}
    \label{eq:LBMX}
    M_X \gtrsim 1.5 \times 10^5 \GeV \;
    \frac{1}{d_{\rm tot}^{\frac{2}{3}}} \,.
\end{eqnarray}
Furthermore, 
the branching ratio 
of the decay channel $X \to \psi_{3/2} + \psi_{3/2}$ is 
\begin{eqnarray}
    \label{eq:B32_MODULI}
    B_{3/2} \equiv \frac{ \Gamma_{3/2} }{ \Gamma_X }
    = \frac{ d_{3/2}^2 }{ 36 \, d_{\rm tot}^2 } \,.
\end{eqnarray}
It is seen that $B_{3/2}$ can be much smaller than unity if $d_{3/2}
\ll 1$ (and/or if $d_{\rm tot} \gg 1$).
Indeed, $B_{3/2}$ should be suppressed enough to avoid cosmological
difficulties, as we will see below.

Now we are at the position to estimate the gravitino abundance.
At the reheating epoch, gravitinos are produced by the decay process
$X \to \psi_{3/2} + \psi_{3/2}$ as discussed in the previous section.  
The yield of the gravitinos produced by this process 
 is estimated as
\begin{eqnarray}
    \label{eq:Y32X0}
    Y_{3/2}^X \simeq \frac{3}{2} B_{3/2} \frac{T_R}{M_X} \,,
\end{eqnarray}
which is defined by the ratio between the gravitino and
entropy densities. Moreover, gravitinos are produced by the thermal scatterings
at the reheating.  We denote this contribution by $Y_{3/2}^{\rm TH}$.
The total yield is then given by 
\begin{eqnarray}
    \label{eq:Y32}
    Y_{3/2} = Y_{3/2}^X + Y_{3/2}^{\rm TH} \,.
\end{eqnarray}
Notice that $Y_{3/2}$ remains constant as the universe expands,
as long as there is no additional entropy production, and we assume it
in the present analysis.

We should mention here that there are potential sources of gravitinos
in addition to the above mentioned ones.   
First, gravitinos would be produced much before the decay of
$X$ such that the abundance remains sizable even after the dilution by the
reheating.  Second, gravitinos would be produced by 
the decay $X \to \wt X + \psi_{3/2}$ where $\wt X$ is the fermionic
partner of the scalar field $X$.  In order to open this decay channel,
the large mass hierarchy between $X$ and $\wt X$ is required 
(see, e.g., Ref.~\cite{Kohri:2004qu}).  In the following 
we will not consider these possible contributions.
Finally, the decay of $X$ generally produces superparticles, {\it i.e.} 
$R$-parity odd particles, followed by cascade decay into  gravitinos.  
In particular, 
this contribution may be important when gravitino is the LSP.  We will come back to this point later.   

From now on, we will derive the cosmological constraints on the  gravitinos
produced at the reheating by the $X$ decay.  Since the constraints
strongly depend on whether gravitino is unstable or stable, we will discuss
each case separately.
\subsection{Unstable gravitino}
%
\begin{figure}[tb]
 \begin{center}
  \includegraphics[scale=1.5]{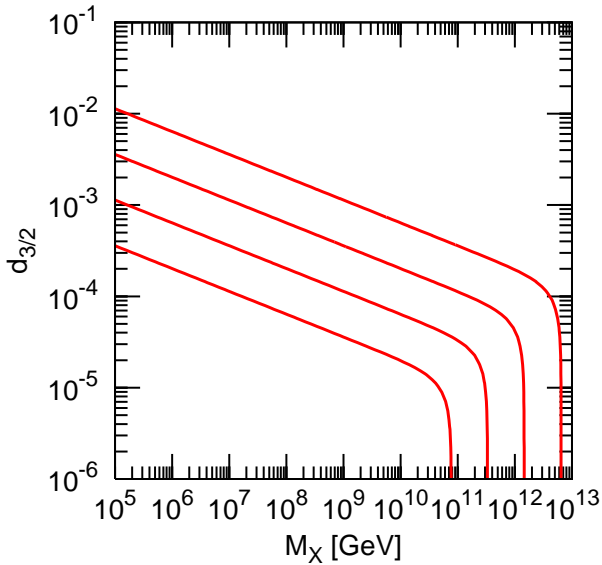}
 \end{center}
 \caption{
   Contour plot of $Y_{3/2}$ in the plane of 
   $M_X$ and $d_{3/2}$ when $\Gamma_X$ is given by
   Eq.~(\ref{eq:GAMX_MODULI}) with $d_{\rm tot}=1$.
   The solid lines correspond to $Y_{3/2} = 10^{-16}$, $10^{-15}$, 
   $10^{-14}$ and $10^{-13}$ from left to right, respectively.
 }
 \label{fig:Y32_MODULIB32}
\end{figure}
%
We first consider the unstable gravitino, especially when the
gravitino mass lies in $M_{3/2} \sim 100$ GeV --$10$ TeV as suggested by
the gravity mediation models of SUSY breaking.  The heavy gravitino
with $M_{3/2} \sim 1$ TeV decays soon after the BBN epoch.  The decay
rate of the gravitino into the MSSM particles is estimated as $\Gamma
\simeq 193/(384 \pi) M_{3/2}^3/ M_P^2$, corresponding to the lifetime
$\tau \simeq 2.4 \times 10^{4} \mbox {sec} \, (1 \TeV
/M_{3/2})^{3/2}$.  In this case, the yield $Y_{3/2}$ is bounded from
above, {\it i.e.} $Y_{3/2} < Y_{3/2}^{\rm BBN}$, in order to keep the
success of the standard scenario of BBN~\cite{Khlopov:1984pf}.
Otherwise, the decay products of the gravitino would change the
abundances of primordial light elements too much and consequently
conflict with the observational data.  The recent
analysis~\cite{Kohri:2005wn} shows that, when the hadronic branching
ratio of the gravitino decay is of order unity, $Y_{3/2}^{\rm BBN} \sim 10^{-16}$
for $M_{3/2} \sim 1$ TeV and $Y_{3/2}^{\rm BBN} \sim
10^{-15}$--$10^{-13}$ for $M_{3/2} \sim 10$ TeV. The constraint
disappears only when the gravitino mass is very heavy: $M_{3/2}
\gtrsim 100$ TeV. We will not consider such a heavy gravitino, as the
region $M_{3/2} \sim 10^2 -10^3 $ TeV is excluded by the overclosure
of the LSPs \cite{Nakamura:2006uc} and the allowed region with
$M_{3/2} \gtrsim 10^3$ TeV is disfavored as the solution to the
naturalness problem inherent to the weak scale.

The yield of the gravitinos produced by the thermal scatterings is
estimated as~\cite{Bolz:2000fu,Kawasaki:2004qu}
\begin{eqnarray}
    \label{eq:Y32TH}
    Y_{3/2}^{\rm TH} \simeq 
    1.1 \times 10^{-12}
    \left( \frac{T_R}{10^{10}\GeV} \right) \,,
\end{eqnarray}
which increases linearly as $T_R$ increases.%
\footnote{We have neglected here the contribution
from the helicity 1/2 components of the gravitino.}
On the other hand, the yield of the gravitinos by the $X$ decay is given by
Eq.~(\ref{eq:Y32X0}).

Consider the case where the $X$ field decays through Planck suppressed 
interaction, in which the total decay rate $\Gamma_X$ is 
given as 
Eq.~(\ref{eq:GAMX_MODULI}) 
with $d_{\rm tot}={\cal O}(1)$. Then
\begin{eqnarray}
    Y_{3/2}^{\rm TH} 
    \eqn{\simeq}
    6.5 \times 10^{-18}  d_{\rm tot} \,
    \left( \frac{M_X}{10^{10}\GeV} \right)^{\frac 32} \,,
    \nonumber \\
    Y_{3/2}^X 
    \eqn{\simeq}
    8.9 \times 10^{-6} \, B_{3/2} d_{\rm tot} \, 
    \left( \frac{M_X}{10^{10}\GeV} \right)^{\frac 12} \,,
\end{eqnarray}
with the total abundance $Y_{3/2}=Y_{3/2}^{X} + Y_{3/2}^{\rm TH}$.  In
this case, $Y_{3/2}$ is determined by $M_X$ and $B_{3/2}$, and
$Y_{3/2}^X \gtrsim Y_{3/2}^{\rm TH}$ for $M_X \lesssim 1.4 \times
10^{10} (B_{3/2}/ 10^{-12} ) \; \GeV $.  In
Fig.~\ref{fig:Y32_MODULIB32} we show the constant contours of
$Y_{3/2}$ in the plane of $M_X$ and $d_{3/2}$.  Here we have set
$d_{\rm tot}=1$.  It is seen that the mass $M_X$ (and hence $T_R$) is
bounded from above by the BBN constraint $Y_{3/2} \simeq Y_{3/2}^{\rm
  TH} < Y_{3/2}^{\rm BBN}$.  Moreover, in order to avoid the
overproduction of the gravitinos by the $X$ decays, the coupling
constant $d_{3/2}$ should satisfy
\begin{eqnarray}
    d_{3/2} \lesssim 2.0 \times 10^{-5} d_{\rm tot}^{- \frac 1 2}  \,
    \left( \frac{M_X}{10^{10}\GeV} \right)^{- \frac 1 4}
    \left( \frac{Y_{3/2}^{\rm BBN}}{10^{-16}} \right)^{\frac 1 2} \,,
\label{eq:constraint-d32}
\end{eqnarray}
or equivalently 
\begin{eqnarray}
    \label{eq:UB_B32_MODULI}
    B_{3/2} \lesssim 1.1 \times 10^{-11} d_{\rm tot}^{-1}
    \left( \frac{M_X}{10^{10}\GeV} \right)^{- \frac 1 2} 
    \left( \frac{Y_{3/2}^{\rm BBN}}{10^{-16}} \right) \,.
\end{eqnarray}

This gives a stringent constraint on the  decay into the
gravitino pair, or equivalently, the $F$-component VEV of the
$X$ field.  In fact, when the total decay rate is controlled by the
Planck suppressed interaction ($d_{\rm tot}={\cal O}(1)$), this
constraint excludes the case of $d_{3/2} = {\cal O}(1)$ as far as the
gravitino mass is $M_{3/2} \sim 100$ GeV--$10$ TeV. Thus, the
cosmological moduli problem cannot be solved by simply raising the
moduli mass~\cite{Endo:2006zj,Nakamura:2006uc}. At the same time, the above
constraint poses a severe restriction to inflation models. In particular,
the modular inflation scenario where one of the moduli fields is used as
the inflaton would face a serious difficulty. 

The bound~(\ref{eq:constraint-d32}) also implies that the scalar decay
survives the BBN constraint if $d_{3/2}$ is small and/or $d_{\rm tot}$
is large.  The former can be achieved by some cancellation among the
contributions in Eq.~(\ref{eq:expression-F_X}), or by the VEV of the
$X$ field $\langle X \rangle$ smaller than the Planck scale. The
latter can be achieved if the $X$ field has stronger interaction to
the MSSM particles than the Planck suppressed one. 

\begin{figure}[tb]
 \begin{center}
  \includegraphics[scale=1.5]{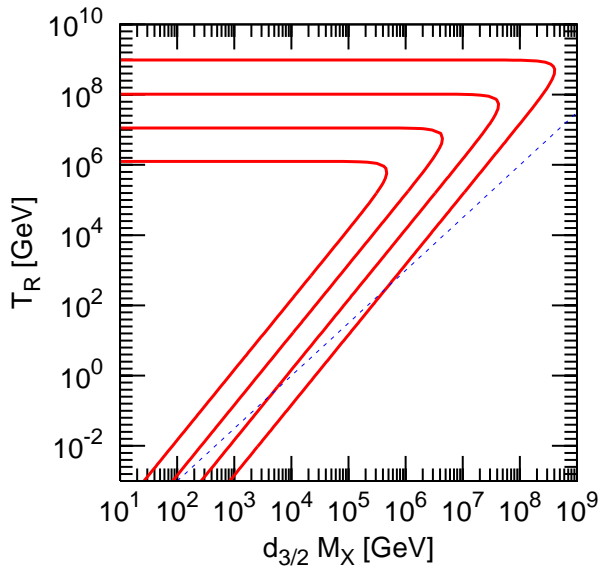}
 \end{center}
 \caption{
   Contour plot of $Y_{3/2}$ in the plane of 
   $d_{3/2} M_X$ and $T_R$.
   The solid lines correspond to $Y_{3/2} = 10^{-16}$, $10^{-15}$, 
   $10^{-14}$ and $10^{-13}$, from inside to outside, respectively.
   The dotted line is the lower bound on $T_R$ from $B_{3/2} \le 1$
   for $d_{3/2} = 10^{-10}$.
 }
 \label{fig:Y32}
\end{figure}

To illustrate such a more general case, it may be convenient to use
$T_R$ and take it as a free parameter.  
In this case, from Eqs.~(\ref{eq:G32}) and (\ref{eq:defTR}) the branching
ratio $B_{3/2}$ is written as
\begin{eqnarray}
    \label{eq:B32}
    B_{3/2} = \frac{d_{3/2}^2}{2 88 \pi}
    \left( \frac{90}{\pi^2 g_\ast (T_R)} \right)^{\frac 1 2}
      \frac{M_X^3}{T_R^2 M_P} \,,
\end{eqnarray}
where $B_{3/2} \le 1$ as it should be.
Then, $Y_{3/2}^{\rm TH}$ is given by Eq.~(\ref{eq:Y32TH}), while
$Y_{3/2}^{X}$ can be written by using Eq.~(\ref{eq:B32}) as
\begin{eqnarray}
    \label{eq:Y32X}
    Y_{3/2}^X \simeq \frac{1}{192 \pi}
    \left( \frac{90}{\pi^2 g_\ast (T_R)} \right)^{\frac 12}
      \frac{d_{3/2}^2 M_X^2}{T_R M_P} \,.
\end{eqnarray}
Quite interestingly, $Y_{3/2}^X$ is inversely proportional to $T_R$,
whereas  $Y_{3/2}^{\rm TH}$ is proportional to $T_R$.
This means that the BBN observation puts not only the upper bound but
also the lower bound on the reheating temperature.  Further, it is
seen that the total yield $Y_{3/2} = Y_{3/2}^X + Y_{3/2}^{\rm TH}$ can
be determined from two parameters, i.e. $d_{3/2} M_{X}$ and $T_R$.
In Fig.~\ref{fig:Y32} we draw the contour plot of $Y_{3/2}$ in the
plane of $d_{3/2} M_X$ and $T_R$.   Here we have fixed
$g_\ast(T_R) = 200$, for simplicity.%
\footnote{
This choice is unsuitable for $T_R \lesssim 1$ TeV, however,
our final results do not change much.}
It is clearly seen that
$Y_{3/2}^{\rm BBN}$ puts the upper bound on $d_{3/2} M_X$ as 
well as $T_R$.  This bound can be found as follows:
In general, $Y_{3/2} = Y_{3/2}^X + Y_{3/2}^{\rm TH}
\ge 2 \sqrt{ Y_{3/2}^X  Y_{3/2}^{\rm TH} } \equiv Y_{3/2}^{\rm MIN}$, 
where the equality holds when $Y_{3/2}^X = Y_{3/2}^{\rm TH}$, i.e.,
$T_R \simeq 1.2 d_{3/2} M_X$ from Eqs.~(\ref{eq:Y32TH}) and
(\ref{eq:Y32X}).  It is obtained that
\begin{eqnarray}
    Y_{3/2}^{\rm MIN} \simeq
    2.6 \times 10^{-22} \, \GeV^{-1} \, d_{3/2} M_X \,,
\end{eqnarray}
and hence $Y_{3/2}^{\rm MIN} < Y_{3/2}^{\rm BBN}$ gives 
the upper bound
\begin{eqnarray}
    \label{eq:UBMX}
    M_X \lesssim 4 \times 10^5 \GeV \,
    \frac{1}{d_{3/2}}
    \left( \frac{Y_{3/2}^{\rm BBN}}{10^{-16}} \right) \,.
\end{eqnarray}
Notice that we have assumed $M_X \gg M_{3/2} \sim 1$ TeV in this case.
Therefore, when $d_{3/2} = {\cal O}(1)$, this gives a sever bound 
on the mass of the scalar field $X$.

We should note that the parameter space in Fig.~\ref{fig:Y32} 
may contradict $B_{3/2} \le 1$.  Indeed, the viable reheating
temperature is
\begin{eqnarray}
    T_R \ge 10^{-5} \GeV \, \frac{1}{d_{3/2}^{\frac 12}}
    \left( \frac{d_{3/2} M_X }{10^4 \GeV} \right)^{\frac 3 2} \,.
\end{eqnarray}
As an example, we also show in Fig.~\ref{fig:Y32} this lower bound on
$T_R$ for $d_{3/2} = 10^{-10}$.  It is thus found that the bound is
insignificant as long as $d_{3/2}$ is sufficiently large.

Finally, when $Y_{3/2} = Y_{3/2}^{\rm MIN}$, the branching ratio 
is $B_{3/2} \simeq 1.8 \times 10^{-4} M_X/M_P$ and then from
Eq.~(\ref{eq:UBMX}) we find
\begin{eqnarray}
    d_{3/2} B_{3/2} \lesssim 2.9 \times 10^{-17}
    \left( \frac{Y_{3/2}^{\rm BBN}}{10^{-16}} \right) \,.
\end{eqnarray}
This again tells that, in order to avoid the overproduction of the gravitinos,
we have to require (i) very small $d_{3/2}$ and/or 
(ii) very small $B_{3/2}$.  The former one may be realized by
a VEV of $X$ much smaller than $M_P$, while the latter one
may demand interactions of $X$ to the MSSM particles, whose strength
is much larger than $ 1/M_P$, to increase $d_{\rm tot}$.

\begin{figure}[tb]
 \begin{center}
  \includegraphics[scale=1.5]{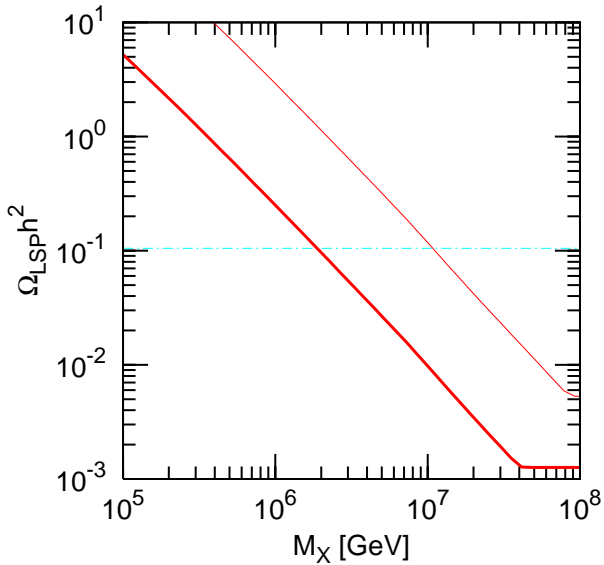}
 \end{center}
 \caption{ Present abundance of the wino LSP $\Omega_{\wt W}h^2$ when
   $\Gamma_X$ is given by Eq.~(\ref{eq:GAMX_MODULI}) with $d_{\rm
     tot}=1$.  The thick and thin solid lines are for
   $M_{\wt W} =100$ and 300 GeV, respectively.
   The horizontal dot-dashed line is the present dark matter 
   density $\Omega_{\rm dm} h^2 \simeq 0.105$.}
 \label{fig:OMwino_MODULI}
\end{figure}
Next, we would like to discuss an additional constraint concerning 
the abundance of the LSPs.
The LSP is stable when the $R$-parity is conserved.  
To be consistent with the current
observation of the cosmic microwave background
radiation~\cite{Spergel:2006hy}, the present LSP abundance should be
\begin{eqnarray}
    \Omega_{\rm LSP} h^2 \le \Omega_{\rm dm} h^2 
    = 0.105^{+0.008}_{-0.014} \,,
\end{eqnarray}
where $h \simeq 0.73$~\cite{Spergel:2006hy} is the present Hubble constant in
the unit of 100km/sec/Mpc.

The LSP is produced by the decays of the gravitinos discussed above.
We then obtain the upper bound on the yield of the gravitinos as
\begin{eqnarray}
    Y_{3/2} < 3.8 \times 10^{-12}
    \left( \frac{ \Omega_{\rm dm} }{ 0.105 } \right)
    \left( \frac{ M_{\rm LSP} }{100 \GeV } \right)^{-1} \,,
\end{eqnarray}
with $M_{\rm LSP}$ being the LSP mass.  We can see that the bound is much
weaker than the constraint from the BBN ($Y_{3/2}^{\rm BBN}$), and it
gives no additional bound.%
\footnote{ This is the case, provided that we consider the gravitino 
 with $M_{3/2}
  \sim 100$GeV--10TeV.  If  the gravitino mass is heavier, this gives
  a stringent constraint as discussed in
  Refs.~\cite{Nakamura:2006uc}.}
In addition, the LSP is produced directly by the $X$ 
decay~\cite{Moroi:1994rs,Kawasaki:1995cy,Moroi:1999zb}.
Here we consider the neutral wino $\wt W$ as the LSP to maximize the 
annihilation cross section leading to the most conservative bound.
The details of estimating the present abundance of $\wt W$ are 
found in Appendix~\ref{sec:AP}.

In Fig.~\ref{fig:OMwino_MODULI} we show $\Omega_{\wt W} h^2$ in terms
of $M_X$ by using Eq.~(\ref{eq:TRMODULI}).  We find that $\Omega_{\wt
  W}h^2$ becomes constant for $M_X \gtrsim 4 \times 10^7$ GeV for
$M_{\wt W} =100$ GeV, while $\Omega_{\wt W}h^2$ becomes larger as
$M_X$ decreases. (See the discussion in Appendix~\ref{sec:AP}).
Therefore, we obtain the lower bound $M_X$ to avoid the overclosure by
the wino LSP, and $M_X \gtrsim 2 \times 10^6$ GeV for
$M_{\wt W} =100$ GeV.  The bound becomes more stringent for larger 
$M_{\wt W}$.  For example, $M_X \gtrsim 10^7$ GeV for $M_{\wt W}=300$ GeV.
On the other hand, when
we take $T_R$ as a free parameter, $\Omega_{\wt W}h^2$ depends on
$M_X$ as well as $T_R$.  As estimated in Appendix~\ref{sec:AP}, the
smaller values of $M_X$ and $T_R$ are excluded.
  This restricts the cosmologically viable
parameters of the field $X$.  Finally, we should notice that the bounds become
severer when the LSP is composed of other neutralino components.

\subsection{Stable gravitino}
We now turn to the case of stable gravitino.  This is the case when
the gravitino is the LSP with exact $R$-parity conservation.  In this
situation, the present abundance of the gravitinos is bounded from
above in order to avoid the overclosure of the
universe~\cite{Pagels:1981ke,Moroi:1993mb}.
For this reason, let us estimate the density parameter of 
the gravitinos, which is given by the present energy density of 
the gravitino divided by the critical density $\rho_{\rm cr}$.
The present abundance of the gravitinos produced by the thermal
scatterings is given by~\cite{Bolz:2000fu}
\begin{eqnarray}
    \label{eq:OM32TH}
    \Omega_{3/2}^{\rm TH} h^2
    \simeq 0.21 
    \left( \frac{T_R}{10^{10}\GeV} \right)
    \left( \frac{M_{3/2}}{100\GeV} \right)^{-1} \,.
\end{eqnarray}
Note that $\Omega_{3/2}^{\rm TH}$ is dominated by the contribution
from the helicity 1/2 components of gravitino and it is inversely
proportional to $M_{3/2}$.
On the other hand,  for the gravitinos from the $X$ decay, 
we find from Eq.~(\ref{eq:Y32X0}) that
\begin{eqnarray}
    \label{eq:OM32X}
    \Omega_{3/2}^X = \frac{ M_{3/2} Y_{3/2}^X }{\rho_{\rm cr}/s_0}
    = \frac{0.027}{h^2} \left( \frac{M_{3/2}}{1 \GeV} \right)
    \left( \frac{Y_{3/2}^X}{10^{-10}} \right) \,,
\end{eqnarray}
where $s_0$ is the present entropy density and $\rho_{\rm cr}/s_0 \simeq
3.6 \times 10^{-9}$ $h^2$ GeV.  
Thus, the total abundance is 
$\Omega_{3/2} = \Omega_{3/2}^X + \Omega_{3/2}^{\rm TH}$.
To avoid the overclosure by gravitinos, we must require
\begin{eqnarray}
    \Omega_{3/2} h^2 \le \Omega_{\rm dm} h^2 \,.
\end{eqnarray}
Especially, when $\Omega_{3/2} = \Omega_{\rm dm}$, the  gravitinos 
constitutes the 
dark matter of the universe.  As seen from Eq.~(\ref{eq:OM32X}), 
the bound on $Y_{3/2}^X$ in the present case 
is much weaker than that from the BBN for
unstable gravitinos.

Furthermore, we have to take into account an additional constraint.
Indeed, the gravitinos produced by the $X$ decay face a constraint
from the cosmic structure formation.  Since the momentum of the
gravitino at the production ($p \simeq M_X/2$) can be much larger than
its mass, its free-streaming at the epoch of the matter-radiation
equality may erase the small scale structures which are observed
today.  This warm dark matter constraint leads to the upper bound on
the present velocity dispersion of the gravitino
$v_0$~\cite{Jedamzik:2005sx}.  The power spectrum inferred from the
Ly-$\alpha$ forest data together with the cosmic microwave background
radiation and galaxy clustering constraints puts severe
limits~\cite{Viel:2005qj,Seljak:2006qw}.  From
Ref.~\cite{Seljak:2006qw} we find approximately $v_0 \lesssim 4 \times
10^{-8}$. On the other hand, the present velocity of gravitinos
produced by the $X$ decay is estimated as
\begin{eqnarray}
    v_0 =
    \frac{1}{2} 
    \left( \frac{ g_{\ast s} (T_0) }{ g_{\ast s}(T_R) } \right)^{\frac
    13}
    \sqrt{ 1 - \frac{4 M_{3/2}^2}{M_X^2}}
    \frac{ T_0 M_X }{ M_{3/2} T_R}  \,,
\end{eqnarray}
where $T_0 = 2.35 \times 10^{-13}$ GeV is the present photon
temperature and $g_{\ast s}(T_0) = 43/11$.  Using Eqs.~(\ref{eq:Y32X0}) 
and (\ref{eq:OM32X}), we find that
\begin{eqnarray}
    \Omega_{3/2}^X 
    \simeq \frac{3 B_{3/2}}{2}
    \frac{ M_{3/2} T_R }{ M_X ( \rho_{\rm cr}/s_0 )}
    \,.
\end{eqnarray}
Therefore, $v_0$ can be written as
\begin{eqnarray}
    v_0 \simeq 1.2 \times 10^{-4}  \, B_{3/2} \,,
\end{eqnarray}
where we  assume that $\Omega_{3/2}^X h^2 = \Omega_{\rm dm} h^2
\simeq 0.105$.  Thus, the warm dark matter constraint on
the gravitinos
is translated to  the upper bound on the branching ratio as%
\footnote{
A similar discussion can be applied for the gravitinos produced by the
thermal scatterings.  Although the
constraint from the structure formation puts the lower bound 
on the gravitino mass,  the interesting mass region which will be
discussed here is far above this bound.}
\begin{eqnarray}
    \label{eq:UB_B32}
    B_{3/2} \lesssim 3 \times 10^{-4} \,.
\end{eqnarray}
It should be noted that this warm dark matter constraint can become
weaken, as the portion of the gravitino dark matter in the whole dark
matter density becomes smaller. Ref.~\cite{Viel:2005qj} gives the
upper bound to $\Omega_{3/2}^X \lesssim 0.12 \; \Omega_{\rm dm}$ to
eliminate the warm dark matter constraint.  Here we shall use this as
a representative bound.

Another potential constraint comes from the BBN, since 
additional energy from the gravitinos by the $X$ decay may increase 
the Hubble expansion rate at $T \sim 1$ MeV too much, which results
in the overproduction of $^4$He.  The bound, however, is rather weak: 
$B_{3/2} \lesssim 0.35$.

\begin{figure}[tb]
 \begin{center}
  \includegraphics[scale=1.5]{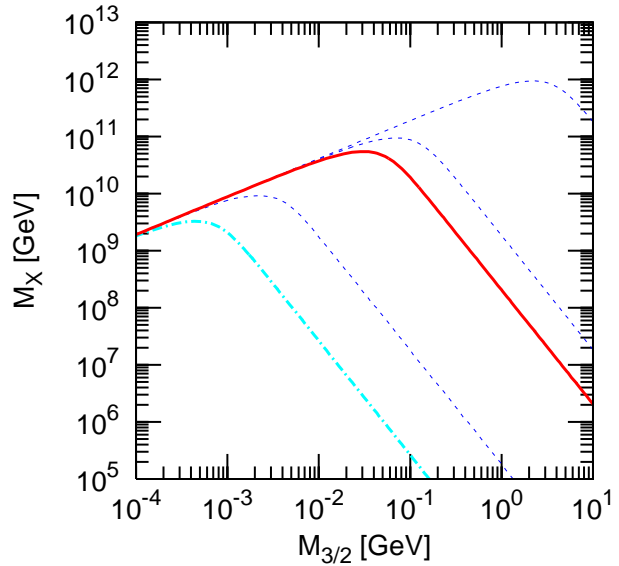}
 \end{center}
 \caption{
   Upper bounds on $M_X$ in terms of $M_{3/2}$, when $\Gamma_X$ is
   given by Eq.~(\ref{eq:GAMX_MODULI}) with $d_{\rm tot}=1$.  The
   dotted lines correspond to the upper bounds from $\Omega_{3/2} h^2
   \le \Omega_{\rm dm} h^2 = 0.105$ for $B_{3/2} = 10^{-2}$, $10^{-4}$
   and $10^{-6}$ from left to right, respectively.  The solid line is
   the bound from $\Omega_{3/2} \le \Omega_{\rm dm}$ for $B_{3/2} = 3
   \times 10^{-4}$, and the gravitino dark matter becomes viable above
   this line.  The dot-dashed line is the upper bound on $M_X$ from
   $\Omega_{3/2}^X \le 0.12 \Omega_{\rm dm}$ avoiding the warm dark
   matter constraint in addition to $\Omega_{3/2} \le \Omega_{\rm
     dm}$, when $B_{3/2}=10^{-2}$.  }
 \label{fig:OM32moduli}
\end{figure}

Let us first consider the case where the $X$ decays through the Planck
suppressed interaction and the total decay rate of $X$ is given by
Eq.~(\ref{eq:GAMX_MODULI}).  In this case, we find from Eqs.~(\ref{eq:OM32TH})
and (\ref{eq:OM32X}) that
\begin{eqnarray}
    \Omega_{3/2}^{\rm TH} h^2 \eqn{\simeq} 
    3.9 \times 10^{-2}  d_{\rm tot} 
    \nonumber 
    \\
    \eqn{}\times
    \left( \frac{M_X}{10^{11} \GeV} \right)^{\frac 32}
    \left( \frac{M_{3/2}}{0.1 \GeV} \right)^{-1} \,,
    \\
    \Omega_{3/2}^{X} h^2 \eqn{\simeq} 
    2.5 \times 10^{-1} d_{\rm tot} 
    \left( \frac{B_{3/2}}{10^{-4}} \right)
    \nonumber \\
    \eqn{}\times
    \left( \frac{M_X}{10^{11} \GeV} \right)^{\frac 12}
    \left( \frac{M_{3/2}}{0.1 \GeV} \right) \,.~~~
\end{eqnarray}
Notice that $T_R$ is determined from $M_X$ as shown in
Eq.~(\ref{eq:TRMODULI}).  In Fig.~\ref{fig:OM32moduli} we show the
contour lines of $\Omega_{3/2} h^2 = 0.105$ in the plane of $M_{3/2}$
and $M_X$ by varying $B_{3/2}$.
It should be noted that $\Omega_{3/2}^{\rm
  TH} \propto M_{3/2}^{-1}$ while $\Omega_{3/2}^{X} \propto M_{3/2}$.  
Therefore, for the heavier gravitino mass region,
$\Omega_{3/2}^{X} \gg \Omega_{3/2}^{\rm TH}$ and the gravitinos produced
by the $X$ decay contribute significantly the present energy 
of the universe.
It is clearly seen that one can escape the overclosure of the 
if the mass $M_X$ (and hence $T_R$) is sufficiently small.  The upper bound on
$M_X$ scales as $M_{3/2}^{\frac 23}$ for $\Omega_{3/2} \simeq
\Omega_{3/2}^{\rm TH}$ with smaller gravitino masses, whereas it
scales as $M_{3/2}^{-2}$ for $\Omega_{3/2} \simeq \Omega_{3/2}^X$ with
heavier gravitino masses.  
When $B_{3/2} \gtrsim 3 \times 10^{-4}$, the upper bound on $M_X$ 
becomes more stringent due to the warm dark matter constraint (see
Fig.~\ref{fig:OM32moduli}).  

We should stress here that the cosmological moduli problem can be
solved in the small gravitino mass region.  We expect for moduli
fields that $B_{3/2} \sim 10^{-2}$ from $d_{3/2} = {\cal O}(1)$ and
$d_{\rm tot} = {\cal O}(1)$.  Even in this case, there indeed exists
the parameter space avoiding cosmological difficulties of gravitinos,
where the gravitino mass is $M_{3/2} \lesssim 0.1$ GeV and the moduli
($X$) mass is $M_X \sim 10^5$--$10^9$ GeV, corresponding to $10^4 \;
\GeV \gtrsim T_R \gtrsim 7$ MeV.  Here the lower bound on $M_X$ comes
from Eq.~(\ref{eq:LBMX}), whereas the upper bound is due to the warm
dark matter constraint.  The required large hierarchy between
$M_{3/2}$ and $M_X$ can be realized in a class of models of moduli
stabilization.  (See, for instance, Ref.~\cite{Kallosh:2004yh}.)

Here we would like to point out that the heavy scalar $X$ decay into a
pair of gravitinos can offer an interesting and alternative window of
dark matter.  When $B_{3/2} \lesssim 3 \times 10^{-4}$, the
free-streaming effect of the produced gravitino is small so that the
gravitino can constitute the dark matter of the universe, {\it i.e.}
$\Omega_{3/2}^{\rm X} = \Omega_{\rm dm}$.  \footnote{ The gravitinos
  produced by the thermal scatterings can be cold dark matter of the
  universe when $M_{3/2} \lesssim 1$ MeV as long as $\Omega_{3/2}^{\rm
    TH} = \Omega_{\rm dm}$~\cite{Moroi:1993mb}.} In
Fig.~\ref{fig:OM32moduli} we present the contour line of $\Omega_{3/2}
h^2 = 0.105$ with $B_{3/2}=3 \times 10^{-4}$ by the solid line.  In
the region above this line, the gravitino from the $X$ decay can
become the viable dark matter as long as $\Omega_{3/2}^X = \Omega_{\rm
  dm}$.

We should mention that the properties of the gravitino dark matter can
be different for different choices of parameters. Namely, the
gravitino becomes the warm dark matter for $B_{3/2} \sim 3 \times
10^{-4}$, and it gets {\it cooler} and eventually becomes the cold
dark matter as the branching ratio decreases.  Furthermore, the mixed
scenario of cold and warm dark matter is possible.  In particular, the
gravitinos produced by $X$ can compose the warm component while those
from the thermal scatterings can compose the cold one.  More precise
observations on the small scale structures of the universe enable us
to test these hypotheses.

The small branching ratio $B_{3/2} \sim 10^{-4}$ indicates from
Eq.~(\ref{eq:B32_MODULI}) that $d_{3/2} \sim 0.1 d_{\rm tot}$.  Though
for the moduli fields a naive expectation will be $d_{3/2} \sim d_{\rm
  tot}$, the suppression of one order of magnitude may also be
possible in some moduli stabilization mechanism. If it is the case, the
decays of the moduli field with mass of ${\cal O}(10^9)$--${\cal
  O}(10^{11})$ GeV will yield the gravitino warm dark matter.

So far, we have assumed that the total decay rate of the $X$ field is
given by Eq.~(\ref{eq:GAMX_MODULI}) with $d_{\rm tot}$ not very for
from unity.  Now we take $\Gamma_X$ or $T_R$ as a free parameter.
Notice that we find from Eq.~(\ref{eq:B32}) that
\begin{eqnarray}
    M_X \eqn{=} 
    \left( \frac{288\pi}{d_{3/2}^2} \right)^{\frac 13}
    \left( \frac{\pi^2 g_\ast (T_R)}{90} \right)^{\frac 16}
    B_{3/2}^{\frac 13} \, T_R^{\frac 23} \, M_P^{\frac 13}  \,,
\end{eqnarray}
which leads to
\begin{eqnarray}
    \label{eq:OM32X1}
    \Omega_{3/2}^X h^2
    \eqn{=}
    4.1 \times 10^{-2} \, 
    \left( \frac{d_{3/2} B_{3/2}}{10^{-6}} \right)^{\frac 23}
    \nonumber \\
    \eqn{}\times
    \left( \frac{T_R}{10^4 \GeV} \right)^{\frac 13}
    \left( \frac{M_{3/2}}{1 \GeV} \right) \,.
\end{eqnarray}
The total abundance $\Omega_{3/2} = \Omega_{3/2}^{X} +
\Omega_{3/2}^{\rm TH}$ is then determined by the two parameters, $T_R$
and $d_{3/2} B_{3/2}$, when the gravitino mass $M_{3/2}$ is given.  

Similar to the case with unstable gravitino, we can obtain 
the upper bound on $d_{3/2} M_X$ to avoid the overclosure by
gravitinos.  Since $\Omega_{3/2} \ge \Omega_{3/2}^{\rm MIN} 
= 2 \sqrt{\Omega_{3/2}^X \Omega_{3/2}^{\rm TH}}$, 
$\Omega_{3/2}^{\rm MIN} h^2 < \Omega_{\rm dm} h^2 \simeq 0.105$ 
results in the bound
\begin{eqnarray}
    M_X \lesssim 5.7 \times 10^9 \GeV \, \frac{1}{d_{3/2}} \,.
\end{eqnarray}
Further, if the branching ratio becomes larger than
Eq.~(\ref{eq:UB_B32}), this upper bound becomes tighter due to the 
warm dark matter constraint.  We can see the bound is much weaker 
than Eq.~(\ref{eq:UBMX}) in the case of unstable gravitino.

\begin{figure}[t]
 \begin{center}
  \includegraphics[scale=1.5]{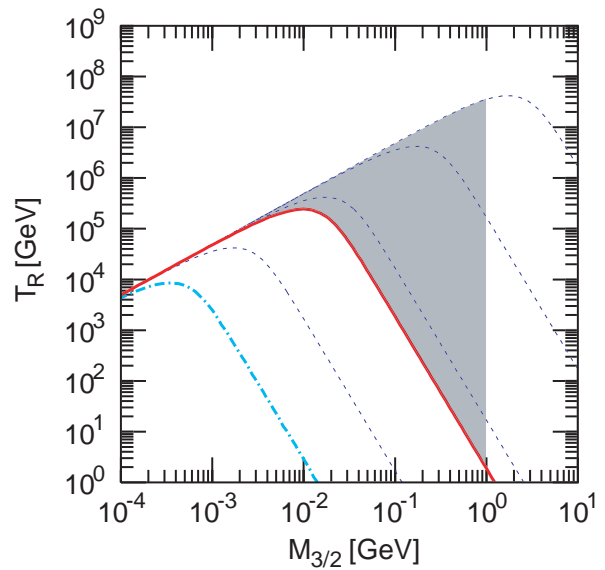}
 \end{center}
 \caption{
   Upper bounds on $T_R$ in terms of $M_{3/2}$.  The dotted lines
   correspond to the upper bounds from $\Omega_{3/2} h^2 \le
   \Omega_{\rm dm} h^2 = 0.105$ for $d_{3/2} B_{3/2} = 10^{-2}$,
   $10^{-4}$, $10^{-6}$ and $10^{-8}$ from left to right,
   respectively.  The solid line is the bound from $\Omega_{3/2} \le
   \Omega_{\rm dm}$ for $d_{3/2} B_{3/2} = 3 \times 10^{-4}$, and the
   gravitino dark matter becomes viable above this line shown as the
   shaded region.  The dot-dashed line is the upper bound on $T_R$
   from $\Omega_{3/2}^X \le 0.12 \Omega_{\rm dm}$ avoiding the warm
   dark matter constraint in addition to $\Omega_{3/2} \le \Omega_{\rm
     dm}$, when $d_{3/2} B_{3/2}=10^{-2}$.  }
 \label{fig:OM32}
\end{figure}
\begin{figure}
 \begin{center}
  \includegraphics[scale=1.5]{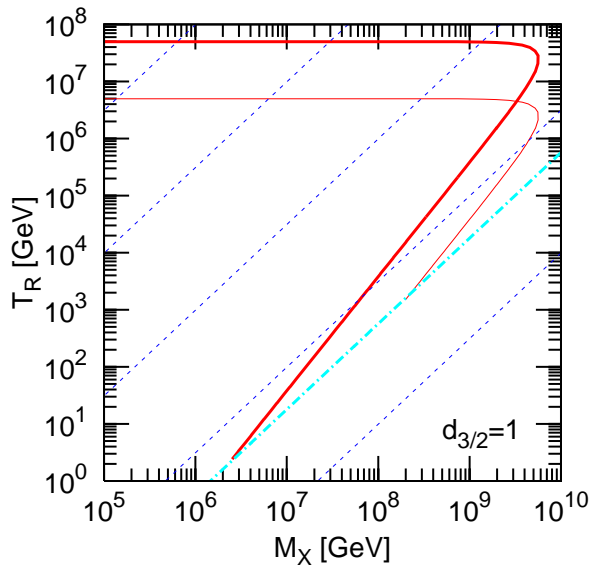}
 \end{center}
 \caption{ Allowed parameters for the gravitino dark matter with
   $M_{3/2}=1$ GeV (the thick solid line) and $M_{3/2}=0.1$ GeV (the
   thin solid line) when $d_{3/2}=1$.  Contour lines of the branching
   ratio are shown by the dotted lines, which correspond to $B_{3/2}=1$,
   $10^{-5}$, $10^{-10}$, $10^{-15}$ and $10^{-20}$ from below,
   respectively.  The upper bound on $B_{3/2}$ from the cosmic
   structure formation is shown by the dot-dashed line.  }
 \label{fig:OM32_d32=1}
\end{figure}
In Fig.~\ref{fig:OM32}, we show the contour plot of $\Omega_{3/2} =
\Omega_{\rm dm}$ in the plane of $M_{3/2}$ and $T_R$ by varying
$d_{3/2} B_{3/2}$.  Notice that the upper bound on
$B_{3/2}$~(\ref{eq:UB_B32}) can be applied even in the general case,
and hence we expect $d_{3/2} B_{3/2} \lesssim 3 \times 10^{-4}$ 
since $d_{3/2} \lesssim {\cal O}(1)$.
In Fig.~\ref{fig:OM32} we also show $\Omega_{3/2}
h^2 \simeq 0.105$ with $d_{3/2} B_{3/2} = 3 \times 10^{-4}$.  
Moreover, the upper bound on $T_R$ from the warm dark matter
constraint is also shown when $d_{3/2} B_{3/2} = 10^{-2}$.
We find that the allowed region is similar to 
that in Fig.~\ref{fig:OM32moduli}, but the dark matter window for the
 gravitinos
produced by the $X$ decay is enlarged.

Further, we show in
Fig.~\ref{fig:OM32_d32=1} the allowed parameter set of the gravitino
dark matter for $d_{3/2} = 1$ by taking the gravitino mass as
$M_{3/2}=1$ GeV and 0.1 GeV.  For smaller $d_{3/2}$, the corresponding
branching ratio becomes smaller as shown in Eq.~(\ref{eq:B32}).  It
can be seen that the gravitino dark matter is possible in a wide range
of parameter space, and an important implication is that $X$
should have the suppressed branching ratio of the decay into a pair of
gravitinos as shown in Eq.~(\ref{eq:UB_B32}).

Finally, we should argue that the NLSP (next-to-LSP) decay into gravitino
restricts the allowed parameter space.
Here let us consider, as an example,  
the case when a (right-handed) scalar tau $\wt \tau$ is the NLSP.
The lifetime of $\wt \tau \to \psi_{3/2} + \tau$ is
\begin{eqnarray}
    \tau_{\wt \tau} 
    \simeq 5.9 \times 10^4 \mbox{sec}
    \left( \frac{M_{3/2}}{1\GeV} \right)^2
    \left( \frac{M_{\wt \tau}}{100 \GeV} \right)^{-5} \,,
\end{eqnarray}
and $\wt \tau$ decays during the BBN period or later for $M_{3/2} \gtrsim
10$ MeV when we take, {\it e.g.}, 
$M_{\wt \tau} = 100$ GeV.  Therefore, when the gravitino
mass is sufficiently small, the $\wt \tau$ NLSP becomes cosmologically
harmless.  On the other hand, when the gravitino mass becomes larger,
the decay products would
dissociate or overproduce the light elements synthesized at the BBN epoch
and would conflict with the observations.  To avoid this, the
number of $\wt \tau$ at the decay time should be small enough.
In Appendix~\ref{sec:AP}, we estimate the abundance of $\wt \tau$,
$Y_{\wt \tau}$ in terms of $M_X$ and $T_R$.

The BBN constraint on $Y_{\wt \tau}$ ($M_{\wt \tau} Y_{\wt \tau}$ more
precisely) strongly depends on how $\wt \tau$ decays into
hadrons~\cite{Kawasaki:2004qu,Kohri:2005wn,Feng:2004zu}.  According to
the hadronic branching ratio $B_h$ in Ref.~\cite{Feng:2004zu} and
also to the BBN constraints for $B_h = 1$ and $10^{-3}$ in
Ref.~\cite{Kawasaki:2004qu}, we find that two regions, $\tau_{\wt
  \tau} \lesssim 10^{2}$ sec and $10^6$ sec $\gtrsim \tau_{\wt \tau}
\gtrsim 10^3$ sec, are cosmologically viable.  
Here we have used $Y_{\wt \tau} \sim 10^{-13}$ by assuming that
the reheating temperature $T_R$ is higher than the freeze-out temperature
of the stau annihilation, $T_F$(see the discussion in Appendix).
Although we cannot find
the relevant constraint for $10^3$ sec $\gtrsim \tau_{\wt \tau}
\gtrsim 10^2$ sec (i.e. $B_h = 10^{-3}$--1) in the literature, we
consider such a window is also allowed from the rough estimate of the
BBN constraints interpolating in the region $B_h = 10^{-3}$--1.  In
this analysis, therefore, we take $\tau_{\wt \tau} \lesssim 10^6$ sec
as a representative bound, which leads to $M_{3/2} \lesssim 4 \GeV$
for $M_{\wt \tau}$.  We should stress here that our final conclusions
leave intact even when the bound on $\tau_{\wt \tau}$ becomes severer.
Since the lifetime of $\wt \tau$ depends on $M_{\wt \tau}^5$, 
the upper bound on $M_{3/2}$ can be enlarged by larger $M_{\wt
  \tau}$.
Finally, this analysis is true for $T_R \gtrsim T_F$. 
On the other hand, for $T_R \lesssim T_F$,
the bound on $M_{3/2}$ becomes stronger or weaker in the low or high
$M_X$ region, respectively.
The bottom line is that 
the BBN constraint on the $\wt \tau$ NLSP decay can be escaped for $M_{3/2} <
{\cal O}(1)$ GeV at least when the reheating temperature is higher
than about $T_F$.  The detail analysis in other
parameter space will be done in future publication.

The upper bound on the gravitino mass of $M_{3/2} \lesssim {\cal
  O}(1)$ GeV implies that the maximal reheating temperature is $T_R
\simeq 5 \times 10^{7}$ GeV.  Such a high reheating temperature is
possible only when $B_{3/2} \lesssim 10^{-6}$ such that $\Omega_{3/2}
\simeq \Omega_{3/2}^{\rm TH}$.  For larger $B_{3/2}$ the upper bound
on $T_R$ is suppressed due to the gravitino production from the $X$
decay.  Note that the upper bound on $T_R$ is directly translated into
the upper bound on $M_X$ through Eq.~(\ref{eq:TRMODULI}).  For
instance, $T_R \lesssim 5 \times 10^{7}$ GeV gives $M_X \lesssim 9
\times 10^{11}$ GeV for $d_{\rm tot}=1$.  (Note that $M_X \gtrsim 1.5
\times 10^5$ GeV to have $T_R \gtrsim 7$ MeV.)

\section{Conclusions and Discussion}
In this paper, we have considered the cosmological implications to the
decay of the general heavy scalar field into the gravitino pair. Here we
would like to summarize what we have obtained in this analysis. 

As was shown in the previous works \cite{Endo:2006zj,Nakamura:2006uc},
the decay amplitude to the gravitino pair is proportional to the VEV
of the auxiliary component $\langle F^X \rangle$ of the heavy field
$X$. We have thus presented the estimate for the VEV of $F^X$ in a
general setting: we have considered the general coupling between $X$
and the fields responsible for the spontaneous supersymmetry breaking,
and also we have considered the case of the explicit supersymmetry
breaking as well. In both cases, we have obtained the same and simple
estimate for this value when only a single $X$ field participates. The
result shows that generally the VEV of $F^X$ is proportional to the
VEV of the $X$ field, and thus the partial decay rate into the
gravitino pair is suppressed when the $X$'s VEV is smaller than the
Planck scale.

We have then considered the various constraints on the gravitino
production at the decay of the heavy scalar field $X$. The relevant
constraints are different whether the gravitino is stable or not, and
so we have discussed the two cases separately. In the unstable
gravitino case, the constraints we have considered are
\begin{enumerate}
\item The BBN constraint on the decay of
  the gravitinos, producing hadronic as well as electromagnetic activities.
\item The constraint on the abundance of the LSPs produced at the
  gravitino decay.
\item The constraint on the  abundance of the LSPs produced 
  directly at the $X$ decay.
\end{enumerate}

It is well-known that the BBN constraint puts the upper bound on the
reheat temperature $T_R$ in order to suppress the gravitino yield
produced by the thermal scatterings~\cite{Khlopov:1984pf}. In the case at
hand, the decay of the $X$ field produces the gravitinos as
well. Since its yield becomes larger when we lower the reheat
temperature, the BBN bound gives the lower bound on $T_R$. We have
identified the allowed range of $T_R$ both in the case where the $X$
field is moduli-like and in the more general case. We have confirmed
that if the gravitino mass lies in 100 GeV--10 TeV range, no allowed region
exists when $S$ is a moduli-like field, namely the field whose the
decay into gravitinos is not suppressed ($d_{3/2}={\cal O}(1)$) and
the decay into other particles is controlled by the Planck suppressed
interaction ($d_{\rm tot}={\cal O}(1)$). On the other hand, for a more
general case, we have found that the viable region of the parameter
space does exist, but is severely constrained, as was shown in
Fig.~\ref{fig:Y32}.

We have seen that the second constraint is weaker than the first one,
provided that the gravitino mass is in the range given above. On the
contrary, the third constraint excludes the case of very low reheat
temperature, as the annihilation processes of the neutralino LSPs are
not very effective there and thus the LSP abundance exceeds the
observational abundance of the dark matter in the universe.

For the stable gravitino 
case, we have discussed the following constraints:
\begin{enumerate}
\item The constraint on the gravitino abundance to avoid the 
overclosure of the universe.
\item The constraint from the warm dark matter,
namely the free streaming of the gravitino produced by the heavy $X$ 
decay is small enough for the gravitino to be a viable warm dark matter.
\item The BBN constraint on the decay of the NLSP particles into 
the gravitinos.
\end{enumerate}

The first constraint is similar to the first one in the unstable
gravitino case, but numerically in the stable gravitino case it is
less severe. This makes a wider region of the parameter space
cosmologically viable. In particular the constraint on the gravitino
abundance allows the moduli-like scalar field when the gravitino mass
is lighter than 1-100 MeV, depending on the mass of the $X$ field.

The second constraint given in the above list also gives a significant
constraint. We have found that it puts the upper bound on the
branching ratio as $B_{3/2} \lesssim 3 \times 10^{-4}$ when the
gravitino warm dark matter is the dominant component of the dark
matter. This constraint disappears if the gravitino dark matter
contribution less than 12\% of the total dark matter density.

The third constraint is also quite stringent, but rather involved. With the 
lack of the complete analysis all through the relevant parameter regions,
we have argued to put the upper bound on the gravitino mass (or the lifetime of
the NLSP) as roughly of order 1 GeV when the NLSP is the stau weighing 100 GeV.

\

In the rest of the paper, we would like to discuss the implications of
our results to the inflationary scenarios. Assuming that there is no
entropy production at a later epoch, our consideration gives a
stringent constraint on the reheating process right after the
inflation. When the inflaton is one of the moduli 
fields~\cite{Blanco-Pillado:2004ns,Blanco-Pillado:2006he,Lalak:2005hr}, and the
oscillating field is in fact moduli-like, then our results severely
constrain the allowed gravitino mass region. In particular for the
unstable gravitino, the analysis in Ref. \cite{Nakamura:2006uc} can
apply, leaving only the very heavy gravitino, {\it e.g.} for the wino LSP 
$M_{3/2} \gtrsim 10^3$ TeV, which is disfavored as
the solution to the naturalness problem on the weak scale. On the
other hand, the case of a light and stable
gravitino becomes cosmologically viable (see Fig. 5). However the low
reheat temperature is required to suppress the gravitino abundance
produced through thermal scattering, which may be inconsistent with a
class of modular inflation with inflaton mass around $10^{10}$
GeV.  It is interesting to note that a new window of the gravitino
warm dark matter opens up in which the gravitino with the mass around
100 MeV, produced by the heavy moduli decay, constitutes the warm dark
matter. Furthermore the reheat temperature is of the order $10^5$ GeV,
which is a natural range we expect with the modular inflaton mass
given above.  The price we have to pay to realize this fascinating
case is a slight suppression of the parameter $d_{3/2}$ by one order
of magnitude, which, we suspect, should be possible within the
framework of modular inflation.

As an important remark, we would like to emphasize that the above
argument to the modular inflation can also apply to the cosmological
moduli problem. In the unstable gravitino case, the cosmological
constraints are too strong to be escaped unless the gravitino mass is
heavier than, say, $10^3$ TeV as was discussed above. On the contrary,
the cosmological moduli problem can be solved when the gravitino is
light and stable, say $M_{3/2} \lesssim 0.1$ GeV.  Such light
gravitinos can be realized in the models of the gauge-mediated
supersymmetry breaking.  Compared to the modular inflation where the
$X$ mass is rather high, it is anticipated that the moduli mass is not
very far from the electroweak scale. Thus somewhat different region in
the parameter space may be favored. For instance, the $X$ mass much
below $10^{10}$ GeV is also fine in this case. Anyway, the solution
requires the large hierarchy between masses $M_{3/2}$ and $M_X$, which
can be realized in a class of models of the moduli
stabilization~\cite{Kallosh:2004yh}.

Let us come back to the implications to the inflationary models. 
Our results given in Section 3 indicate that, to make the inflaton
decay cosmologically viable in a wider range of the parameter space, a
smaller branching ratio of the decay into the gravitino pair (a
smaller $d_{3/2}$) and a larger total decay rate (a larger $d_{\rm
  tot}$) is favored. In a simple class of the chaotic inflation, the
Lagrangian is invariant under a $Z_2$ discrete symmetry, a reflection
of the field variable $X \to -X$. Thus at the minimum of the vacuum,
$F_X$ will vanish and thus the decay of the inflaton into the
gravitino pair does not take place and so the model does not suffer
from the gravitino production problem. There are other inflationary
models in which the parameter $d_{3/2}$ becomes much smaller than
unity, by realizing the VEV of $X$ in an intermediate scale lower than
the Planck scale. Examples include a new inflation model and also a
hybrid inflation. Whether these models are really viable or not
require detailed case study.  A consideration can be found in 
Ref.~\cite{Kawasaki:2006gs}. 

Finally to make $d_{\rm tot}$ large, one can construct an inflation
model where the inflaton interacts to other particles with
renormalizable couplings.  An example is given as follows.  Suppose an
inflaton $X$ couples to a pair of right-handed neutrinos in the
superpotential as $W = (f_N/2) X N^c N^c$, where $f_N$ is a coupling
constant and the Majorana mass of $N^c$ is given by $M_N = f_N \langle
X \rangle$.  Now, we set $\langle X \rangle = 10^{15}$ GeV.
\footnote{Such a VEV of the $X$ field may be realized in the
  supersymmetric model of the new and hybrid
  inflation~\cite{Asaka:1999jb}.}  
In this case, we find $B_{3/2} \sim 10^{-9}$ from the decay rate in
Eq.~(\ref{eq:GAMX_MODULI}).  The inflaton decay rate of $\Phi \to N^c
+ N^c$ is $\Gamma_{N^c} = M_N^2 M_X/(32 \pi) /\langle X \rangle^2$,
which becomes much larger in some parameter region.
 For example, $B_{3/2} \sim 10^{-12}$
for $M_X = 10^{10}$ GeV and $M_N = 10^9$ GeV by taking
$d_{3/2} = \langle X \rangle/M_P$.
Due to this fact, the cosmological constraints
become weaker.

\acknowledgments

We would like to thank T. Moroi and A. Yotsuyanagi for fruitful
discussisons.  The work was partially supported by the grants-in-aid
from the Ministry of Education, Science, Sports, and Culture of Japan,
No.~16081202 and No.~17340062.
\appendix*
\section{}
\label{sec:AP}
In this appendix, we briefly explain an estimation of the abundance
of the lightest superparticle in the MSSM, which is denoted by $A$.
In the text, we consider $A$ as the neutral wino for unstable
gravitinos, while $A$ as the stau for stable gravitinos.

\begin{figure}
 \begin{center}
  \includegraphics[scale=1.5]{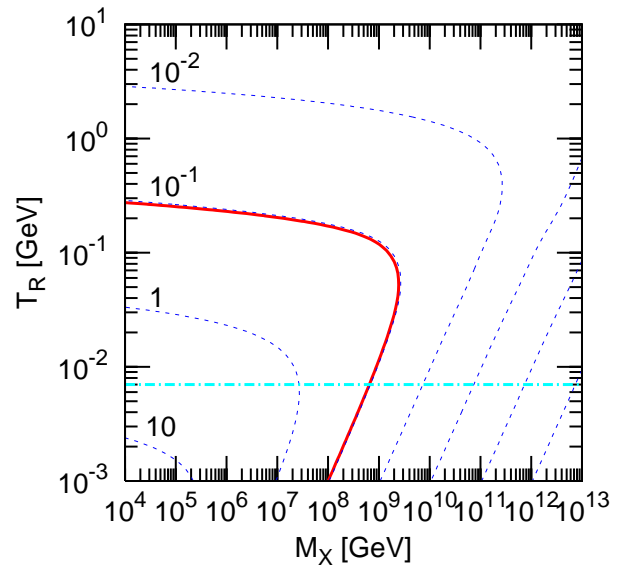}
 \end{center}
 \caption{
   Contour plot of the present abundance of the wino LSP
   $\Omega_{\wt W} h^2$ for $M_{\wt W}=100$ GeV.  The dotted lines are
   $\Omega_{\wt W} h^2 = 10$, 1, \dots, $10^{-5}$ from left to right,
   respectively.  
   For $T_R \gtrsim 10$ GeV,
   $\Omega_{\wt W} h^2 \simeq 1.3 \times 10^{-3}$.
   The solid line shows $\Omega_{\wt W} h^2
   = 0.105$, and the left-bottom region is excluded
   by the overclosure by $\wt W$.  The horizontal dot-dashed line
   represents the lower bound on $T_R (\gtrsim 7$MeV).}
 \label{fig:OMwino}
\end{figure}
%
\begin{figure}
 \begin{center}
  \includegraphics[scale=1.5]{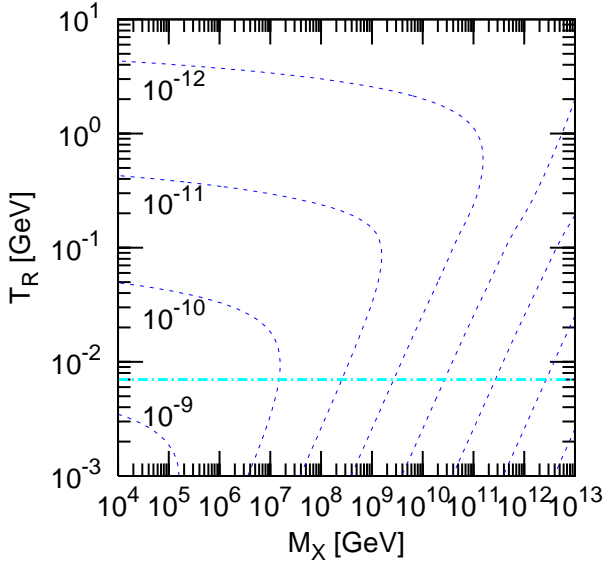}
 \end{center}
 \caption{
   Contour plot of $Y_{\wt \tau}$ in the $M_X$-$T_R$ plane 
   for $M_{\wt \tau} = 100$ GeV.
   The dotted lines correspond to $Y_{\wt \tau}$ = $10^{-9}$--$10^{-16}$
   from left to right.  For $T_R \gtrsim 10$ GeV, 
   $Y_{\wt \tau} \simeq 1.2 \times 10^{-13}$.
   The horizontal dot-dashed line
   represents the lower bound on $T_R (\gtrsim 7$MeV).
 }
 \label{fig:Ystau}
\end{figure}
The decay of $X$ and also the thermal scatterings produce $A$, and
then the number density of $A$, $n_A$, can be found by solving the
following coupled equations (see, for example,
Ref.~\cite{Gelmini:2006pw} and the references therein);
\begin{eqnarray}
    \label{eq:ap1}
    && \dot \rho_X + 3 H \rho_X = - \Gamma_X \rho_X  \,,
    \\
    \label{eq:ap2}
    && \dot \rho_R + 4 H \rho_R = + \Gamma_X \rho_X  \,,
    \\
    \label{eq:ap3}
    && \dot n_A + 3 H n_A = 
    \langle \sigma v \rangle 
    \left( n_{\rm eq}^2 - n_A^2 \right) + 
    \frac{B_A \, \Gamma_X}{M_X} \rho_X \,, ~~
\end{eqnarray}
where $\rho_X$ and $\rho_R$ are the energy densities of the $X$ field
and radiations, respectively, and the dot denotes a time derivative.
$M_X$ and $\Gamma_X$ are the mass and total decay rate of $X$.  $B_A$
denotes the effective number of $A$ per a $X$ decay.  $\langle \sigma
v \rangle$ are the annihilation cross section which is thermally
averaged, and $n_{\rm eq}$ is the equilibrium value of the number
density of $A$.  In these equations, $H$ is the Hubble expansion rate
which is given by $H^2 = ( \rho_X + \rho_R )/(3 M_P^2)$.  The cosmic
temperature is found from $\rho_R$.  Here we have neglected the
co-annihilation effects on $n_A$, and the contribution to $\rho_R$
from $A$ since it can be negligible, and also assumed the rapid
thermalization of $A$ just after the production~\cite{Kawasaki:1995cy,Hisano:2000dz}.

In solving these equations, we assume that the energy of the universe
is dominated by $X$ for $H > \Gamma_X$, and we take the initial
condition such that the maximal temperature of the dilute plasma
(for $H > \Gamma_X$) becomes sufficiently high to keep 
$A$ in the thermal equilibrium.   The decay rate $\Gamma_X$
is parameterized by the reheating temperature $T_R$ as
\begin{eqnarray}
    \Gamma_X = \left( \frac{ \pi^2 g_\ast (T_R) }{90} \right)^{ \frac
    1 2} \frac{T_R^2}{M_P} \,.
\end{eqnarray}

The abundance of $A$ is determined from two parameters $T_R$ and
$B_A/M_X$,  in addition to $\langle \sigma v \rangle$ and $M_A$.
For simplicity, we shall take $B_A = 0.5$ in
this analysis, and the results for other values of $B_A$ can be found 
by the rescaling of $M_X$ as long as $T_R$ is considered as a
free parameter.   Then, $T_R$ and $M_X$ control the abundance of 
$A$, {\it i.e.} the yield of $A$, $Y_A$, after $A$ decouples from the
thermal bath.

When $T_R \gtrsim T_F$ ($T_F$ is the freeze-out temperature of $A$ and
$T_F \sim M_A/20$), $Y_A$ does not depend on $M_X$.  This is because
$A$ is still in the equilibrium after the decay of $X$ completes, and
its abundance is evaluated from $\langle \sigma v \rangle$ and $M_A$
as usual.  On the other hand, when $T_R \lesssim T_F$, $Y_A$ strongly
depends on $T_R$ and $M_X$.  When $M_X$ is sufficiently small, $A$ is
produced so abundantly by the $X$ decays due to the source term in
Eq.~(\ref{eq:ap3}) that $Y_A$ is determined by its annihilation effect
at $H \sim \Gamma_X$.  In this case, $Y_A$ is inversely proportional
to $T_R$.  On the other hand, when $M_X$ is sufficiently large,
the annihilation at $H \sim \Gamma_X$ becomes insignificant
and $Y_A$ is determined from the contributions from the production 
by thermal scatterings in the dilute plasma and also from the $X$
decay at $H \sim \Gamma_X$.   In this case, we find that 
$Y_A \propto T_R$.

Now we present our numerical results.  First, we consider the case
when $A$ is the neutral wino $\wt W$ which is the stable LSP.  
In this case, we use~\cite{Moroi:1999zb}
\begin{eqnarray}
    \langle \sigma v \rangle
    = \frac{ g_W^4 ( 1-x_W )^{\frac 32} }{ 2 \pi ( 2 - x_W )^{ 2} }
    \frac{1}{M_{\wt W}^2} \,,
\end{eqnarray}
where $g_W$ is the weak gauge coupling and $x_W = M_W^2/M_{\wt W}^2$.
In Fig.~\ref{fig:OMwino} we show the contour plot of the present
abundance $\Omega_{\wt W} h^2 = M_{\wt W} Y_{\wt W} h^2 /(\rho_{\rm
  cr}/s_0)$ by taking $M_{\wt W} = 100$ GeV.
For $T_R \gtrsim 10$ GeV, $\Omega_{\wt W} h^2$ takes
a value of $1.3 \times 10^{-3}$.

When $A$ is the stau $\wt \tau$ which is the NLSP, we
use~\cite{Asaka:2000zh}
\begin{eqnarray}
    \langle \sigma v \rangle
    = \frac{ 4 \pi \alpha_{em}^2 }{M_{\wt \tau}^2} \,.
\end{eqnarray}
In Fig.~\ref{fig:Ystau} we show the contour plot of 
the yield $Y_{\wt \tau}$ by taking $M_{\wt \tau} = 100$ GeV.
For $T_R \gtrsim 10$ GeV, $Y_{\wt \tau}$ takes
a value of $1.9 \times 10^{-13}$.

\printfigures

%
\end{document}